# Random temporal laser speckles for robust measurement of sub-microsecond photoluminescence decay

**J. JUNEK,**[1,3] **L. ONDIČ,**[2] **K. ŽÍDEK**[1,*]

[1]*Regional Centre for Special Optics and Optoelectronic Systems (TOPTEC), Institute of Plasma Physics, Czech Academy of Science v.v.i., Za Slovankou 1782/3, 182 00 Prague 8, Czech Republic*
[2]*Institute of Physics, Czech Academy of Sciences v.v.i., Cukrovarnická 10, 162 53, Prague 6, Czech Republic*
[3]*Technical University in Liberec, Faculty of Mechatronics, Informatics and Interdisciplinary Studies Studentská 1402/2, 461 17 Liberec, Czech Republic*
*\* Author to whom correspondence should be addressed. Email: zidek@ipp.cas.cz*

**Abstract:** Time-resolved photoluminescence (PL) is commonly used to track dynamics in a broad range of materials. Thus, the search for simplification of the acquisition of PL kinetics attracts continuous attention. This paper presents a new robust and straightforward approach to the measurement of PL decay, which is based on randomly fluctuating excitation intensity. The random excitation waveform is attained by using laser speckles generated on a rotating diffuser. Owing to this, the presented technique is able to utilize any coherent excitation source without the necessity to generate short pulses or to controllably modulate the light. PL decay can be computationally reconstructed from the Fourier image of the PL trace. The paper demonstrates the performance of the method, which is able to acquire sub-microsecond dynamics as the impulse response function reaches 300 ns. The reconstructed PL decays were compared to streak camera measurements to verify the method. Finally, potential limitations and applications of the technique are discussed.

## 1. Introduction

Photoluminescence (PL) spectra and dynamics provide a vast amount of information about the emitting material – revealing energy levels of charge carriers [1,2], resolving processes governing the excited energy routes [3,4], even resolving lifetimes of the processes [5]. All the information can be extracted in a non-contact experiment, even in opaque samples. Therefore, PL time-resolved spectroscopy counts among the most used characterization methods in the fields of chemical physics [6,7], biochemistry [8], and material sciences in general.

All reported methods for time-resolved PL measurements require an excitation source able to provide us with a short pulse or to be controllably modulated [9-11]. In general, all the methods can be divided into two groups, which measure PL kinetics in the time and the frequency domains. Methods working in the time domain need a source producing correspondingly short pulses. The pulses excite PL, which is either detected by a speedy detection system, is spatially swept (streak camera), or the PL signal is gated (by up-conversion technique, iCCD). Measurements of PL decay in the frequency domain demand using a controllably modulated light intensity, for instance, an acousto-optic modulator or a modulated laser. In order to capture PL decay stretched over different timescales, the intensity modulation has to be facilitated over a broad range of frequencies.

In this article, we present a method that, in contrast can use any source of coherent light to measure PL dynamics on the microsecond and sub-microsecond timescale. The core of this method lies in the excitation of the measured sample with a randomly fluctuating intensity of light. This can be achieved by transmitting a coherent light source through a rapidly varying scattering element creating fluctuating speckle patterns, which we denote as temporal speckles. The field of temporal speckles is cut with an aperture, providing us with a random signal for sample excitation. Such a random excitation signal features a broad range of frequencies, which can be used to reconstruct back the PL decay from the measured fluctuations of the PL intensity. We demonstrate that this technique can be used to attain PL decay with sub-microsecond temporal resolution without the need for a pulsed excitation. We also prove that the attained decay is in agreement with conventional approaches to PL decay measurement.

Moreover, as we show later, this method is remarkably robust against various experimental conditions such as signal delay or offset. Owing to its simplicity, no elaborate setup alignment or calibration is needed.

## 2. Methodology and experimental setup

### 2.1 Principles of the method

For the sake of brevity, we will hereafter denote the method as RATS (Random Temporal Speckles). The goal of the RATS method is to excite the tested sample with temporal speckles, i.e., randomly fluctuating intensity of light. To reconstruct the PL decay $I_{decay}$ we need to acquire the excitation signal $I_{Exc}$ and the PL signal $I_{PL}$. We can illustrate the RATS concept on simulated data depicted in Figure 1, A-B. The excitation signal was attained by simulating the generation of laser speckles from a rotating diffuser via Fraunhofer diffraction (far-field speckle pattern) [12]. Following the real experimental setup, which is described later, the resulting random pattern was cut with an aperture and the total intensity of light within the aperture is plotted in Figure 1A as a random signal for the excitation $I_{Exc}$. The photoluminescence signal $I_{PL}$ plotted in Figure 1B was calculated via the convolution of $I_{Exc}$ and $I_{decay}$:

$$I_{PL} = I_{Exc} * I_{decay} \tag{1}$$

It is worth noting that Eq. (1) holds only for PL intensity which is linearly proportional to the excitation intensity, as we discuss later.

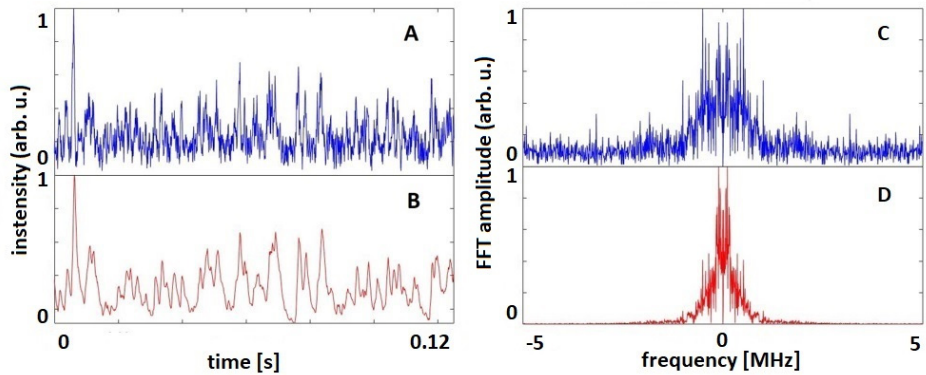

Figure 1. Simulated intensity of the fluctuating temporal speckles (panel A) which are used to calculate a PL waveform (panel B). We assume a single-exponential PL decay with $\tau = 1$ ms. Fourier transform of intensity of fluctuating temporal speckles (panel C) and Fourier transform of PL waveform (panel D).

For the sake of clarity, the PL decay lifetime in Figure 1 was chosen to be mono-exponential with the lifetime in the millisecond range ($\tau = 1$ ms), so that the difference between the excitation and the PL waveform is apparent both in the time domain and in the Fourier space. The Fourier transformations of $I_{Exc}$ and $I_{decay}$ are depicted in Figure 1C and D, respectively.

By using a fluctuating random signal for excitation, we get a wide range of frequencies in the Fourier space. This is important because the $I_{decay}$ can be determined on the timescale corresponding to the highest available frequency. The $I_{decay}$ can be simply calculated via the convolution theorem:

$$I_{decay} = Re\left\{\mathcal{F}^{-1}\left[\frac{\mathcal{F}(I_{PL}).conj(\mathcal{F}(I_{Exc}))}{\mathcal{F}(I_{Exc}).conj(\mathcal{F}(I_{PL}))}\right]\right\} \tag{2}$$

We, therefore, attain the full PL decay dynamics $I_{decay}$, analogously to the other methods. The decay can be subsequently fitted with a set of exponential functions to determine the characteristic PL lifetimes.

Naturally, since the temporal speckle pattern is formed by a rotating diffuser, excitation and PL waveforms repeat with every period of rotation. Nevertheless, for the purpose of the PL decay retrieval, we do not need to determine the frequency of the diffusor and its timing, i.e., phase. This is owing to the fact that the shift in the "zero time" between the PL and excitation data will cause the PL decay to be multiplied with a constant complex number $e^{(-i\varphi)}$, where the phase $\varphi$ will scale with the timing difference. Such a problem can be easily eliminated.

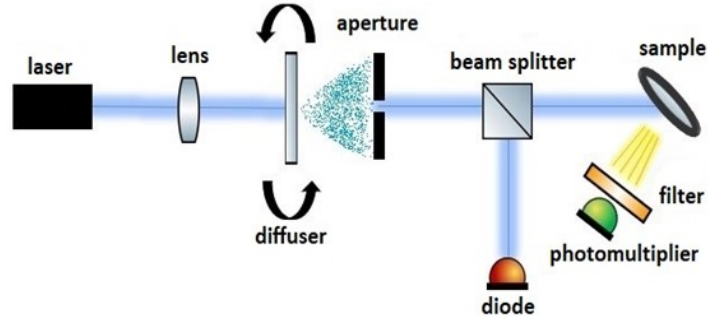

Figure 2. Scheme of the experimental setup used.

## 2.2 Optical setup

A scheme of the optical setup used is presented in Figure 2. In the setup, we used a cw laser (Sapphire, Coherent) at the wavelength of 488 nm, intensity 6 mW. But it should be stressed that one can use any coherent light source which is suitable for excitation of a sample. The laser beam was focused onto a diffusor with a lens with a focal length of 30 mm. Under those conditions, we generated speckle patterns in the so-called far-field regime, which means that the generated intensity pattern does not change with distance. The diffusor (ground glass, the average grain size of 1.52 µm) was mounted on a Mitsumi DC motor, which rotated up to the frequency of 65 Hz.

The field of speckles was cut with an iris aperture at the distance of 110 mm from the diffusor. As a result, we attained a randomly fluctuating intensity of light behind the aperture, which was used to excite the sample of interest. The best results were achieved when the aperture was set to the same size as the mean size of speckles. A larger size of the aperture decreases the contrast of the temporal speckles; a small size reduces the excitation light intensity.

A small fraction of the excitation intensity was divided by a thin BK7 window and directed onto a biased Si photodiode (Thorlabs, 35 ns rise time) to acquire the excitation signal. The remaining part was used to excite a sample. Samples were excited by a fluctuating excitation intensity with the typical mean intensity of 5 µW. The PL signal was detected with a Hamamatsu photomultiplier (PMT) module type H10721-20 (rise time 0.6 ns). The module was placed close to the sample so that no collecting lens needed to be used. A cut-off color filter (OG 515) was placed in front of the PMT to shield out the excitation wavelength. Both the excitation and the PL signals were amplified by an SRS amplifier model SR445A and read out by a TiePie Handyscope HS3 USB oscilloscope.

In order to compare the results of the RATS method with a commonly used technique, we measured the same testing samples on the RATS setup and on a Hamamatsu StreakScope C10627-11 streak camera coupled with a spectrograph. In the streak-camera measurements, we excited the sample with 515 nm pulses featuring a low excitation energy of 15 nJ/pulse and 190 fs pulse length. The spectrally-resolved PL dynamics were attained with the central detection wavelength of 650 nm.

## 3. Results and discussion

### 3.1 Instrument response function

A crucial parameter of time-resolved spectroscopy is the attainable temporal resolution of a method. This is characterized by the impulse response function (IRF) of the method. The IRF can be determined as a reconstruction of scattered excitation light, or by measuring a sample, where PL decays very rapidly compared to the expected IRF width. We measured IRF by using PL from Rhodamine 6G solution, which has got PL lifetime in units of nanoseconds [13], and evaluated its full width half maximum (FWHM) for each experimental parameter. The IRF width is affected by several factors, which we discuss in the following paragraphs.

Analogously to standard methods, the presented method can be limited by the available bandwidth of the used photodetectors and amplifier. Since we used detection systems and electronics with a bandwidth exceeding 300 MHz, this fact did not limit the performance of the RATS setup.

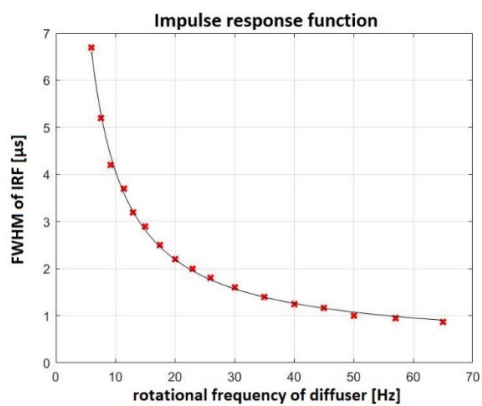

Figure 3. Experimentally measured impulse response function width (FWHM) for different rotation frequencies of diffuser (red crosses) fitted with a reciprocal function (black line). Laser wavelength 488 nm, diffuser diameter 100 nm, grain size 1.52 µm.

Secondly, the detected signal in the RATS experiment has to be sampled on a level where no aliasing occurs. Aliasing can cause severe distortion of the reconstructed PL decay. Therefore, it is essential to adjust the sampling rate according to the highest available frequency in the temporal speckle fluctuation or to use a low-pass frequency filter before the signal sampling.

Finally, it transpired that the main limiting factor in the RATS method case is the rate of temporal speckle fluctuation. The width of the IRF is inversely proportional to the maximum frequency present in the temporal speckle waveform. The frequency of speckles is given by the peripheral speed of the diffusor, which is inversely proportional to the grain size of the diffusor.

To test this fact, we evaluated IRF for an increasing frequency of the rotating diffuser, which was 100 mm in diameter, featuring the mean grain size of 1.52 μm. Since the frequency of the temporal speckles is proportional to the peripheral speed of the diffusor, the IRF becomes inversely proportional to the diffusor rotational frequency, as one can also observe in Figure 3. For a diffusor frequency reaching 65 Hz, the FWHM of the IRF decreased to 870 ns. Hence, the presented setup allowed us to measure PL dynamics with the microsecond and sub-microsecond resolution.

### 3.2 Testing measurements

To test the presented method, we used an orange color filter (SCHOTT, OG 565, glass matrix containing $CdS_xSe_{1-x}$ nanoparticles), which absorbs all wavelengths below 550 nm and can be therefore excited with the used laser (488 nm). Another testing sample was a layer of nanoporous silicon, which was prepared by electrochemical etching of Si wafer in hydrofluoric acid and ethanol solution. Details of the porous silicon preparation process and its optical properties can be found in previous publications, see [2, 14].

Figure 4 summarizes the detected random signals (top panels), their Fourier representation (middle panels), and the determined PL decay for both samples (black crosses in bottom panels). The PL decay of the OG 565 filter was fitted with a sum of two exponentials convoluted with a Gaussian IRF (FWHM 1.2 μs), which corresponds well to the used frequency of the diffuser of 55 Hz (see Figure 4A, left panel). The PL fitted decay of the OG 565 filter (red line) was $Ae^{t/\tau_1} + Be^{t/\tau_2}$, where A = 0.14, B = 0.006, $\tau_1$ = 0.6 μs (sub-IRF), $\tau_2$ = 4.8 μs.

Nanoporous silicon PL features a broad range of lifetimes, as it decays via the so-called stretched exponential function $A.exp[-(t/\tau)^\beta]$ [15]. The measurement of this sample was done for the frequency of the diffuser of 23 Hz (IRF FWHM of 1.9 μs). The determined PL decay (see Figure 4B, right panel) was fitted with the stretched exponential function convoluted with Gaussian IRF (red line). The attained fitted parameters of the PL decay were $\tau$ = 0.7 μs, β = 0.35.

The performance of the RATS method was verified by measuring the PL decay of the two testing samples using a streak camera [16]. We picked the streak camera, as it is a commonly used device with a PL resolution very well below the RATS setup. The streak camera spectrograms were acquired on several timescales for each testing sample. Figure 5A shows a selected spectrogram of the orange filter, in which the vertical axis represents the time axis, whereas the horizontal axis represents various wavelengths.

The attained PL decay curves are plotted in Figure 5B and 5C, where the red lines correspond to the streak camera curves, and the black lines were measured with the RATS method. The PL decays are in perfect agreement with a negligible difference arising due to the fact that the PMT used in the RATS setup and the streak camera feature different spectral sensitivities. Since PL decay depends on PL wavelength (as can be clearly seen in Fig. 5), the total decay can vary for distinct sensitivities.

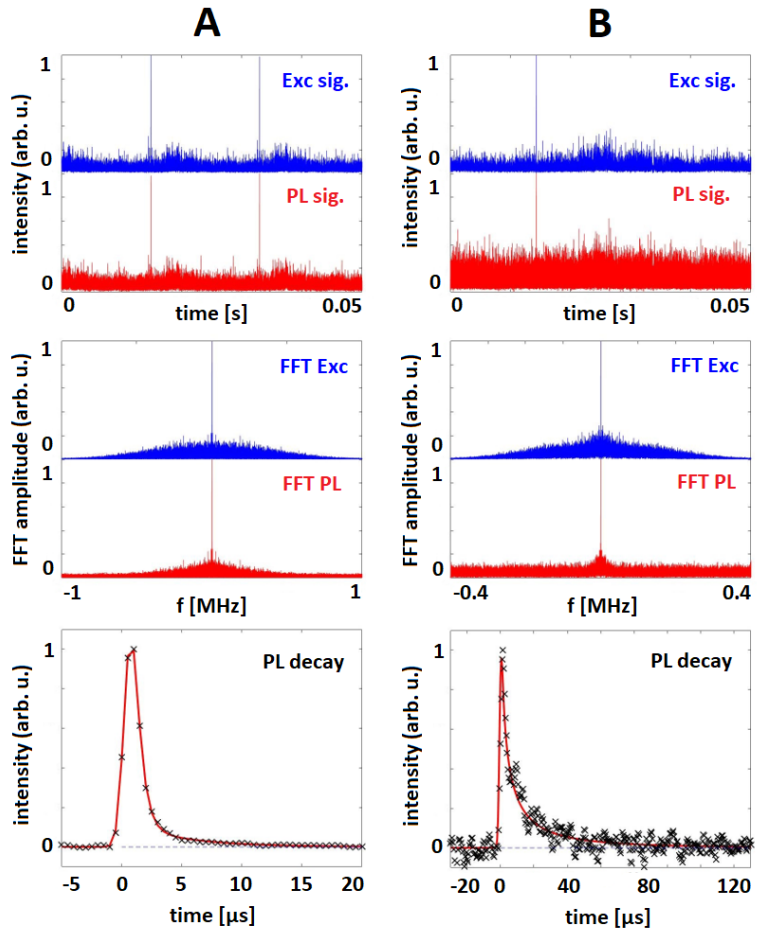

Fig. 4 Results of the RATS method for the orange filter OG565 (A) and nanoporous Si (B). Top panels: measured excitation and PL waveforms; middle panels: Fourier amplitudes of the signals; bottom panels: reconstructed PL kinetics (black crosses) fitted with PL decay convoluted with Gaussian IRF (red lines). λexc = 488 nm. IRF width: OG565: 1.2 µs; nanoporous Si: 1.9 µs.

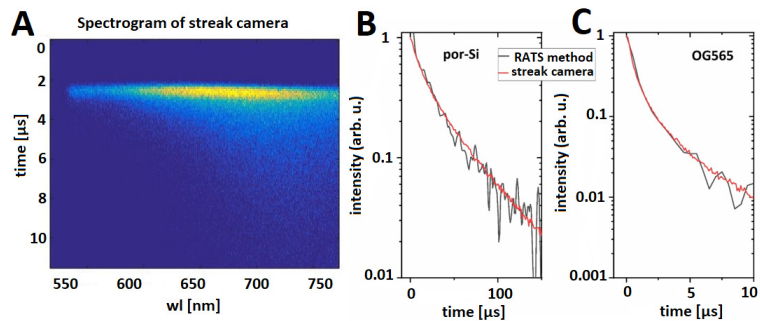

Figure 5. Streak camera spectrogram of the orange filter sample (A). Excitation wavelength: 515 nm; energy 15 nJ/pulse; 190 fs pulse length. Comparison of PL decays of nanoporous silicon (B) and OG 565 filter (C) acquired by RATS method (black curves) and integration of the streak camera data (red lines). For the sake of comparison, the data were normalized.

## 4. Method potential and limitations

The presented method has several advantages compared to its standard counterparts. Firstly, as we stressed previously, the excitation source can be any coherent light source without the necessity to be modulated controllably. Instability of the source is not an issue since it will only positively contribute to the random fluctuations.

Moreover, the method is very robust against signal offsets both in the temporal and detector background sense. A temporal offset between the PL and the excitation intensity will lead to a constant offset in complex phase in the reconstructed decay, i.e., the resulting curve will be the actual decay curve multiplied by $e^{-i\varphi}$. This can be corrected easily. At the same time, the background offset of the signals will manifest only on the zero-frequency edge of the Fourier transform and can again be avoided by removing the low frequencies from the decay reconstruction.

We demonstrated that the method could be used for measurements on the (sub)microsecond scale. The presented method shares with the other methods for measurement of PL decay in the temporal domain the need for a fast detection system and electronics. Nevertheless, such devices are widely available on the market, providing bandwidth of several hundreds of MHz. Therefore, the real limiting factor in our case lies in the properties of the diffusor, i.e., its peripheral speed and its grain size. In principle, we are not limited to any border values. Nevertheless, the size and rotational speed of the diffuser are certainly restricted by the necessity to accommodate the setup in a laboratory and to maintain a safe operation of the setup.

The principal limitation of the method lies in the fact that the intensity of the PL emission must be linearly proportional to the excitation intensity. Under this condition, the measured PL waveform follows equation (1) and the convolution theorem can be used. Nevertheless, since the method relies on using a weak excitation intensity, this condition is satisfied for the majority of materials. Moreover, the method can be extended to be used for more general PL properties, namely, by changing the approach to PL decay reconstruction, where PL properties and lifetimes would be fitted to agree with the measured data.

The presented method can be simplified even more by pre-calibrating the $I_{Exc}$ waveform. For a fixed position of the diffused beam and diffusor center of rotation, the waveform will remain the same and does not need to be measured for each sample. Therefore, the entire optical setup can become even simpler by avoiding using a beam splitter and a reference photodiode.

## 5. Conclusions

We presented a new approach to measuring PL decay, which can be used as a simple means of material characterization. The RATS (RAndom Temporal Speckles) method generates sample excitation by using a rotating diffuser without any need for a dedicated light source. With that approach, the method overcomes demands on advanced optical equipment and becomes a low-cost option for PL decay measurement.

By using this entirely different approach to the excitation signal, we can benefit from several advantages. Namely, no special requirements are put on the excitation source and its stability, and the method uses a low-cost setup enabling rapid data acquisition. Moreover, the random character of the excitation signal allows us to bring in advanced methods for computational reconstruction of PL decay, which will be addressed in our future work.

It should be mentioned that while writing the article, the diffusor and its peripheral speed were optimized and an IRF of 300 ns was achieved. However, the updated setup was not used for the above-described measurements; therefore, the results are not presented as such.

**Funding**

The Czech Academy of Sciences (ERC-CZ/AV-B, project Random-phase Ultrafast Spectroscopy); the Grant Agency of the Czech Republic (Project 17-26284Y); the Ministry of Education, Youth and Sports ("Partnership for Excellence in Superprecise Optics," Reg. No. CZ.02.1.01/0.0/0.0/16_026/0008390). LO acknowledges support from the GACR (Project 19-14523S).

**Acknowledgment**

We gratefully acknowledge Jakub Junek and Jakub Nečásek for their help with improving the diffuser rotation speed.

**Disclosures**

The authors declare no conflicts of interest.